\documentclass[prd,twocolumn,groupedaddress,showpacs,nofootinbib]{revtex4}
\usepackage{graphicx}
\usepackage{dcolumn}
\usepackage{amssymb}
\usepackage{mathrsfs}
\usepackage{amsmath}
\usepackage{epsfig}
\usepackage[dvips]{color}
\usepackage{hhline}

\begin{document}

\title{Baryon asymmetry from left-right phase transition}

\author{Pei-Hong Gu}
\email{peihong.gu@sjtu.edu.cn}

\affiliation{School of Physics and Astronomy, Shanghai Jiao Tong University, 800 Dongchuan Road, Shanghai 200240, China}

\begin{abstract}

We extend the standard model fermions by a mirror copy to realize a left-right symmetry. During a strongly first order phase transition of the spontaneous left-right symmetry breaking, the CP-violating reflections of the mirror fermions off the mirror Higgs bubbles can generate a mirror lepton asymmetry and an equal mirror baryon asymmetry. We then can obtain an ordinary baryon asymmetry through the mirror fermion decays where a dark matter scalar plays an essential role. Benefitted from a parity symmetry for solving the strong CP problem, the cosmic baryon asymmetry can be well described by the ordinary lepton mass matrices up to an overall factor. In this scenario, the Dirac CP phase in the Majorana neutrino mass matrix can provide a unique source for the required CP violation. Furthermore, the Higgs triplet for type-II seesaw as well as the first generation of mirror charged fermions can be allowed at the TeV scale.

\end{abstract}


\maketitle

\section{Introduction}

In order to understand the cosmic baryon asymmetry \cite{pdg2018}, we need a dynamical baryogenesis mechanism. This requires a CPT-invariant theory of particle interactions should satisfy the Sakharov conditions \cite{sakharov1967}: (i) baryon number nonconservation, (ii) C and CP violation, (iii) departure from equilibrium. The standard model (SM) fulfils all of these conditions and then accommodates an electroweak baryogenesis mechanism \cite{krs1985,ckn1993}. However, the SM electroweak baryogenesis can only give a baryon asymmetry far far below the observed value because the Jarlskog determinant \cite{jarlskog1985} suppresses the Kobayashi-Maskawa (KM) CP violation \cite{km1973} in the quark sector, meanwhile, the Higgs boson mass exceeds the bound for a strongly first order phase transition of the electroweak symmetry breaking \cite{shaposhnikov1987}.

We need supplement the SM by new ingredients to realize a successful baryogenesis mechanism. One of the most attractive schemes is to consider a leptogenesis \cite{fy1986} scenario, where some non-SM heavy particles decay to generate a lepton asymmetry through their Yukawa \cite{fy1986}, scalar \cite{mz1992,crv1996,ms1998} or other interactions \cite{gu2012,zhuridov2016}, subsequently, the sphaleron processes partially transfer the produced lepton asymmetry to a baryon asymmetry. In some cases, these heavy particles can be motivated by the famous seesaw \cite{minkowski1977,yanagida1979,grs1979,ms1980} mechanism for generating the small neutrino masses. Thus the leptogenesis in seesaw context can unify the origin of the baryon asymmetry and the neutrino mass. In this conventional leptogenesis scenario, we do not know much about the new fields. Consequently we cannot get an exact relation between the cosmic baryon asymmetry and the neutrino mass matrix. For example, we can expect a successful leptogenesis in some seesaw models even if the neutrino mass matrix does not contain any CP phases \cite{di2001}. Alternatively, in some mirror left-right \cite{ps1974,mp1975,mp1975-2,ms1975} symmetric models, where every ordinary fermion including the SM fermions has a mirror partner \cite{gu2012,bm1989,bm1990,bcs1991,cgnr2013,gu2014,gh2016,abbas2016,abbas2017,abbas2017-2,abbas2017-3,gu2017}, the leptogenesis can be realized by the decays of the mirror neutrinos into the mirror charged fermions and then the decays of the mirror charged fermions into their ordinary partners with a dark matter scalar \cite{gu2012,gu2014,gu2017}. Thanks to the parity symmetry for solving the strong CP problem, this unconventional leptogenesis can have a distinct dependence on the neutrino mass matrix.

It is possible to connect the baryon asymmetry and the neutrino mass in other ways. An interesting attempt is to consider the electroweak baryogenesis by the interactions involving the neutrinos \cite{ckn1990}. Although this attempt failed in some simple models with heavy neutral fermions \cite{ckn1990,hr1997}, it could succeed in a late neutrino mass framework \cite{hmp2005}. We can also consider a dark electroweak baryogenesis scenario \cite{gu2017-2} where the Yukawa couplings for the neutrino mass generation help to simultaneously produce the baryonic and dark matter and hence solve the baryonic-dark-matter coincidence problem and predict the dark matter mass. In this scenario, if we further impose a proper mirror symmetry between the ordinary and dark sectors, both the baryon asymmetry and the dark matter-antimatter asymmetry can be well described by the lepton mass matrices up to an overall factor, in particular, the Dirac CP phase in the Majorana neutrino mass matrix will be the unique source for the required CP violation.

In this paper we shall apply the basic idea of electroweak baryogenesis in a mirror left-right symmetric model to explain the cosmic baryon asymmetry. For this purpose, we shall introduce a mirror copy of the SM fermions to construct the left-right symmetric framework. We shall further assume a strongly first order phase transition of the left-right symmetry breaking. The CP-violating reflections of the mirror fermions off the mirror Higgs bubbles then can generate a mirror lepton asymmetry and an equal mirror baryon asymmetry. The decays of the mirror fermions into their ordinary partners and the dark matter scalar then can lead to an ordinary baryon asymmetry. In the presence of a parity symmetry motivated by solving the strong CP problem, the cosmic baryon asymmetry can be parametrized by the ordinary lepton mass matrices up to an overall factor. Remarkably, the Dirac CP phase in the Majorana neutrino mass matrix is the unique source for the required CP violation. Furthermore, the Higgs triplet for type-II seesaw as well as the first generation of mirror charged fermions could be verified at the colliders since they are allowed at the TeV scale.

\section{The model}

The scalars and fermions are classified in three parts: an ordinary sector (OS), a mirror sector (MS) and a crossing sector (CS), i.e.
\begin{eqnarray}
\label{os}
\!\!\!\!\!\!\!\!\!\!\!\!\!\!\!\!&&\begin{array}{lll}\textrm{OS}:&\phi_{1}^{}(1,2,1,-1)(+,+)\,,&\phi_{2}^{}(1,1,2,-1)(+,+)\,,\\
&\Delta(1,3,1,+2)(+,+)\,,&q^{}_{L}(3,2,1,+\frac{1}{3})(-,+)\,,\\
& d^{}_{R}(3,1,1,-\frac{2}{3})(-,+)\,,&u^{}_{R}(3,1,1,+\frac{4}{3})(-,+)\,,\\
&l^{}_{L}(1,2,1,-1)(-,+)\,,&e^{}_{R}(1,1,1,-2)(-,+)\,;
\end{array}\nonumber\\
\label{ms}
\!\!\!\!\!\!\!\!\!\!\!\!\!\!\!\!&&\begin{array}{lll}\textrm{MS}:&\phi'^{}_{1}(1,1,2,-1)(+,+)\,,&\phi'^{}_{2}(1,1,2,-1)(+,+)\,,\\
&\Delta'(1,1,3,+2)(+,+)\,,&q'^{}_{R}(3,1,2,+\frac{1}{3})(+,-)\,,\\
 & d'^{}_{L}(3,1,1,-\frac{2}{3})(+,-)\,,&u'^{}_{L}(3,1,1,+\frac{4}{3})(+,-)\,,\\
&l'^{}_{R}(1,1,2,-1)(+,-)\,,&e'^{}_{L}(1,1,1,-2)(+,-)\,;
\end{array}\nonumber\\
\label{cs}
\!\!\!\!\!\!\!\!\!\!\!\!\!\!\!\!&&\begin{array}{lll}\textrm{CS}:&\sigma(1,1,1,0)(-,-)\,,&\chi(1,1,1,0)(+i,+i)\,.\end{array}\end{eqnarray}
Here the first brackets following the fields describe the transformation under an $SU(3)_c^{}\times SU(2)_L^{}\times SU(2)_R^{} \times U(1)^{}_{X}$ gauge symmetry, while the second brackets are a $Z_4^{}\times Z_4^{}$ discrete symmetry.

It is easy to see the model can respect a discrete parity symmetry. Accordingly, the $SU(2)_{L,R}^{}$ gauge couplings $g_{L,R}^{}$ are identical, i.e. $g_{L}^{}=g_{R}^{}=g$, meanwhile, the $U(1)_X^{}$ gauge coupling $g_X^{}$ is determined, i.e. $g_X^{}=g g'/\sqrt{g^2_{}-g'^2_{}}$ with $g'$ being the SM $U(1)^{}_Y$ gauge coupling. Furthermore, the allowed Yukawa interactions should be
\begin{eqnarray}
\label{yukawa}
\!\!\!\!\mathcal{L}_Y^{}\!\!&=&\!\!-y_d^{}(\bar{q}_L^{}\tilde{\phi}^{}_{1}d_R^{}\!+\!\bar{q}_R^{}\tilde{\phi}'^{}_{1}d'^{}_L)\!-\!y_u^{}(\bar{q}_L^{}\phi^{}_{2}u_R^{}\!+\!\bar{q}'^{}_R\phi'^{}_{2}u'^{}_L)\nonumber\\
\!\!\!\!\!\!&&\!\!-y_e^{}(\bar{l}_L^{}\tilde{\phi}^{}_{1}e_R^{}\!+\!\bar{l}_R^{}\tilde{\phi}'^{}_{1}e'^{}_L)\!-\!\frac{1}{2}y_\nu^{}(\bar{l}_L^c i \tau_2^{} \Delta l_L^{}\!+\!\bar{l}'^c_R i \tau_2^{} \Delta' l'^{}_R)\nonumber\\
\!\!\!\!\!\!&&\!\!-\!f_d^{}\sigma\bar{d}^{}_R d'^{}_L \!-\!f_u^{}\sigma\bar{u}^{}_R u'^{}_L\!-\!f_e^{}\sigma \bar{e}^{}_R e'^{}_L+\textrm{H.c.}\nonumber\\
\!\!\!\!\!\!&&\!\!\textrm{with}~~f_{d,u,e}^{}=f_{d,u,e}^{\dagger}\,,~~y_\nu^{}=y_\nu^T\,.
\end{eqnarray}
As for the scalar potential, it can contain the following cubic terms,
\begin{eqnarray}
\label{potential}
V\supset \rho \sigma \chi^2_{}+\mu^{}_{ij} \phi^T_{i} i\tau_2^{}\Delta \phi^{}_j +\mu'^{}_{ij} \phi'^T_{i} i\tau_2^{} \Delta' \phi'^{}_j +\textrm{H.c.}\,.
\end{eqnarray}
In the above we have assumed the fermions can only couple to one $[SU(2)]$-doublet scalar. This can be achieved by a softly broken global or discrete symmetry which, however, is not explicitly shown for simplicity.

The $Z^{}_4\times Z_4^{}$ symmetry will not be broken at any scales. This means the crossing scalars $\sigma$ and $\chi$ should not obtain any vacuum expectation values (VEVs). As for the other ordinary and mirror scalars, their VEVs are denoted by
\begin{eqnarray}
v_{1,2}^{}=\langle \phi_{1,2}^{}\rangle\,,~v_{\Delta}^{}=\langle \Delta\rangle\,;~v'^{}_{1,2}=\langle \phi'^{}_{1,2}\rangle\,,~v'^{}_{\Delta}=\langle \Delta'\rangle\,.
\end{eqnarray}
We then expect the following linear combinations of the Higgs doublets,
\begin{eqnarray}
\varphi &=& \frac{v_1^{}\phi^{}_1+v_2^{}\phi^{}_2}{v}~~\textrm{with}~~\langle \varphi\rangle=\sqrt{v_1^2+v_2^2}=v\,,\nonumber\\
\varphi' &=& \frac{v'^{}_1\phi'^{}_1+v'^{}_2\phi'^{}_2}{v'}~~\textrm{with}~~\langle \varphi'\rangle=\sqrt{v'^{2}_1+v'^{2}_2}=v'\,,
\end{eqnarray}
to respectively drive the left-right and electroweak gauge symmetry breaking at two hierarchical scales. For example, we expect $v'^{}_{2}>v'^{}_{1}\gg v^{}_{2}>v^{}_{1}$. This can be achieved by spontaneously or softly breaking the discrete parity symmetry. Furthermore, if the ordinary Higgs triplet $\Delta$ is much heavier than the electroweak scale $v$, its VEV $v_{\Delta}^{}$ can be highly suppressed in a natural way because of the so-called type-II seesaw mechanism \cite{mw1980,sv1980,cl1980,lsw1981,ms1981}. Alternatively, a TeV-scale Higgs triplet $\Delta$ can acquire a tiny VEV $v_{\Delta}^{}$ if its cubic couplings $\mu_{ij}^{}$ to the Higgs doublets $\phi_{1,2}^{}$ are small enough. In this case, the type-II seesaw can be verified at the colliders such as the LHC \cite{fhhlw2008}. Actually, a small coupling $\mu_{ij}^{}$ may be understood by an additional seesaw scheme \cite{ghsz2009,gm2017}, where the lepton number is allowed to be spontaneously broken near the electroweak scale. The fashion is similar for the mirror Higgs triplet $\Delta'$.

After the above symmetry breaking, the ordinary and mirror fermions can obtain their Dirac or Majorana mass matrices through the Yukawa couplings (\ref{yukawa}), i.e. 
\begin{eqnarray}
\!\!\!\!\!\!\!\!\!\!\!\!\!\!\!\!&&m^{}_d=y^{}_d v^{}_1  \propto m'^{}_d=y^{}_d v'^{}_1\,,~~m^{}_u=y^{}_u v^{}_2 \propto m'^{}_u=y^{}_u v'^{}_2\,;\nonumber\\
\!\!\!\!\!\!\!\!\!\!\!\!\!\!\!\!&&m^{}_e=y^{}_e v^{}_1 \propto m'^{}_e=y^{}_e v'^{}_1\,,~~m^{}_\nu=y^{}_\nu v^{}_\Delta \propto m'^{}_\nu=y^{}_\nu v'^{}_\Delta\,.
\end{eqnarray}
Remarkably the above quark mass matrices can solve the strong CP problem without introducing an axion
\cite{bm1989,bm1990,bcs1991}. Furthermore, the leptonic Yukawa couplings,
\begin{eqnarray}
y_{e}^{}&\equiv& \hat{y}_e^{}=\{\bar{y}_{e}^{},\bar{y}_{\mu}^{},\bar{y}_{\tau}^{}\}\,,\nonumber\\
y_{\nu}^{}&\equiv& U_{\textrm{PMNS}}^{}\hat{y}_\nu^{}U^T_{\textrm{PMNS}}=U_{\textrm{PMNS}}^{}\{\bar{y}_{1}^{},\bar{y}_{2}^{},\bar{y}_{3}^{}\}U^T_{\textrm{PMNS}}\,,~~
\end{eqnarray}  
can be sizable, i.e.
\begin{eqnarray}
\bar{y}_\tau^{}<\frac{\sqrt{4\pi} m_\tau^{}}{m_b^{}}\,,~~\bar{y}_{e,\mu}^{}=\frac{\bar{y}_\tau^{}m_{e,\mu}^{}}{m_\tau^{}}\,,~~\bar{y}_{1,2,3}^{}<\sqrt{4\pi}\,.
\end{eqnarray}
Here $U_{\textrm{PMNS}}^{}=U\textrm{diag}\{e^{\alpha_1^{}/2}_{},e^{\alpha_2^{}/2}_{},1\}$ is the leptonic mixing matrix and $U$ contains a Dirac CP phase,
\begin{eqnarray}
\label{pmns}
\!\!\!\!\!\!U\!\!\!&=&\!\!\!\left[\!\!\begin{array}{ccl}
\!\!c_{12}^{}c_{13}^{}& \!\!s_{12}^{}c_{13}^{}&\!\!  s_{13}^{}e^{-i\delta}_{}\\
[2mm] \!\!-\!s_{12}^{}c_{23}^{}\!-\!c_{12}^{}s_{23}^{}s_{13}^{}e^{i\delta}_{}
&\!\!~~c_{12}^{}c_{23}^{}\!-\!s_{12}^{}s_{23}^{}s_{13}^{}e^{i\delta}_{}
&\!\! s_{23}^{}c_{13}^{}\\
[2mm]\!\! ~~s_{12}^{}s_{23}^{}\!-\!c_{12}^{}c_{23}^{}s_{13}^{}e^{i\delta}_{}
& \!\!-\!c_{12}^{}s_{23}^{}\!-\!s_{12}^{}c_{23}^{}s_{13}^{}e^{i\delta}_{}
& \!\!c_{23}^{}c_{13}^{}
\end{array}\!\!\right]\!\!.\nonumber\\
\!\!\!\!\!\!&&\!\!\!\end{eqnarray}

The crossing scalar $\sigma$ is assumed heavier than the third generation of mirror charged fermions. We may expect this heavy scalar to drive an inflation. The other crossing scalar $\chi$ is lighter than the first generation of mirror charged fermions so that it can keep stable to serve as a dark mater particle. This dark matter scalar can mostly annihilate into the SM fields through the Higgs portal interaction \cite{ht2016}. Alternatively, it can mostly annihilate into the TeV-scale Higgs triplet for the type-II seesaw. Such leptophilic dark matter scalar could have some interesting implications on the dark matter searches \cite{los2017}. At the same time, the scattering of the dark matter scalar off the ordinary nucleons can arrive at a testable level through the exchange of the SM Higgs boson  \cite{ht2016,los2017}.

\section{Mirror lepton and baryon number}

Since the left-right symmetry has not been experimentally observed yet, it should be broken before the electroweak symmetry breaking. In addition, the phase transition during the left-right symmetry breaking is assumed strongly first order. These requirements can be achieved by properly choosing the parameters in the scalar potential. For simplicity the details will not be studied here.

During the left-right phase transition, the bubbles of the true ground state of the mirror Higgs scalar $\varphi'$ will nucleate and expand until they fill the universe. The left-right symmetry is unbroken outside the bubbles whereas broken inside the bubbles. Accordingly, the $SU(2)^{}_R$ sphaleron reactions can keep very fast outside the bubbles though they are highly suppressed inside the bubbles. As a mirror Higgs bubble expands, the mirror fermions from the unbroken phase will be reflected off the bubble wall back into the unbroken phase. If the CP is not conserved, we can expect a difference between the reflection probabilities for the mirror fermions and anti-fermions with a given chirality. Due to the CPT conservation and the unitary invariance, the asymmetry between the reflected mirror left-handed fermions and anti-fermions should be opposite to the asymmetry between the reflected mirror right-handed fermions and anti-fermions. Furthermore, the thermal masses and hence the momenta perpendicular to the bubble wall as well as the distributions are different for the mirror left- and right-handed fermions outside the bubbles. The reflected mirror fermions can diffuse and participate in the $SU(2)_R^{}$ sphaleron reactions before the bubble wall catches up. At the time the bubble wall catches up the reflected mirror quarks(leptons), the $SU(2)_R^{}$ sphaleron reactions have partially transferred a net baryon(lepton) number inside the bubbles, other than an opposite baryon(lepton) number outside the bubbles, to a lepton(baryon) number. The final baryon(lepton) number thus should be a sum of the unaffected baryon(lepton) number outside the bubbles and the survival baryon(lepton) number inside the bubbles. This means the baryon asymmetry stored in the mirror quarks must equal the lepton asymmetry stored in the mirror leptons.

The above scenario is just an application of the SM electroweak baryogenseis mechanism in our mirror sector. We now check the lepton number from the reflection of the mirror leptons and anti-leptons off the mirror Higgs bubbles \cite{hs1995},
\begin{eqnarray}
\label{lnumber}
\!\!\!\!n^{r}_{L'}\!\simeq \!\int\! \frac{d\omega}{2\pi}n_0^{}(\omega)\left[1-n_0^{}(\omega)\right]\frac{\Delta k \cdot v_W^{}}{T}\Delta(\omega)+\mathcal{O}(v^2_W)\,.
\end{eqnarray}
Here $\omega$ is the energy, $n_0^{}(\omega)=1/(e^{\omega/T}_{}+1) $ is the Fermi-Dirac distribution, $\Delta k$ is the difference between the right and left-handed mirror lepton momenta perpendicular to the bubble wall, $\Delta(\omega)$ is the reflection asymmetry between the mirror leptons and anti-leptons, and $v_W^{}$ is the advancing wall velocity. The reflected mirror leptons and anti-leptons can diffuse before the bubble wall catches up. The typical distance from the advancing bubble wall to the reflected mirror leptons and anti-leptons is $\sqrt{D^{}_{L'}t}-v_W^{}t$ with $D^{}_{L'}$ being a diffusion constant \cite{fs1993,fs1994}. We have known in the SM, the strong interactions dominate the quark diffusion constant $D_{B'}^{} \sim 6/T $ while the weak interactions dominate the lepton diffusion constant $D_{L'}^{}\sim 100/T$ \cite{jpt1996,jpt1996-2}. In our model, the Yukawa couplings involving the mirror leptons can be larger than the strong coupling and hence can dominate the lepton diffusion. So we can simply estimate the lepton diffusion constant $D_{L'}^{}\sim c^{}_{L'}/T$ with $1<  c_{L'}^{}\lesssim 6$. Within the time $t_D^{}\sim D_{L'}^{}/v^2_W$, the $SU(2)_R^{}$ sphalerons can partially convert the lepton number $n^{r}_{L'}$ to a baryon number $n^{}_{B'}$. The remnant lepton number outside the bubbles and the lepton number inside the bubbles can give a net lepton number $n_{L'}^{}$. By solving the diffusion equations, the baryon and lepton numbers inside the expanded bubbles should be \cite{fs1993,fs1994},
\begin{eqnarray}
\label{blnumber}
\frac{n_{L'}^{}}{s}=\frac{n^{}_{B'}}{s}\sim-\frac{9 \Gamma'^{}_{\textrm{sph}}}{T^3_{}}\frac{D_{L'}^{}}{v_W^2}\frac{n^{r}_{L'}}{s}=-\frac{3^3_{}g^8_R \kappa  c_{L'}^{}}{2^7_{}\pi^4_{}v_W^2}\frac{n^{r}_{L'}}{s}\,,
\end{eqnarray}
where $\Gamma'^{}_{\textrm{sph}}=6 \kappa \left[g^2_R/(4\pi) \right]^4_{}T^4_{}$ is the $SU(2)_R^{}$ sphaleron rate per volume with $\kappa \simeq 20$ being a coefficient \cite{bmr2000}, while $s=2\pi^2_{}g_\ast^{}T^{}_{}/45$ is the one-dimensional entropy density with $g_\ast^{}\simeq 215.5$ being the relativistic degrees of freedom.

We now consider the mirror symmetry in Eq. (\ref{yukawa}) to quantitatively analyse the mirror lepton and baryon numbers (\ref{blnumber}). The thermal masses of the related quasi-particles can be denoted by $\Sigma=\Omega- \gamma^0_{}(2 i  \gamma)$ with \cite{weldon1982,weldon1982-2,jpt1996,jpt1996-2,dn1995}, 
\begin{eqnarray}
\Omega_{l'^{}_R}^{2}\!&=&\!\frac{T^2_{}}{8}\left(\frac{3}{4}g^{2}_{R}+\frac{1}{4}g^2_{X}+\frac{1}{2 } \hat{y}^2_e +\frac{3}{4}U^\ast_{}\hat{y}^{2}_\nu U^T_{}\right)\,,\nonumber\\
\gamma_{l'^{}_R}^{}\!&\sim &\!\frac{T}{32\pi}\left(9g^{2}_{R}+6g^2_{X}+\frac{1}{2}\hat{y}^2_e + \frac{3}{4} U^\ast_{}\hat{y}^{2}_\nu U^T_{} \right),\nonumber\\
\Omega_{e'^{}_L}^{2}\!&=&\!\frac{T^2_{}}{8}\left(g^2_X+ \hat{y}^2_e \right)\,,~~\gamma_{e'^{}_L}^{}\!\sim \frac{T}{32\pi}\left(12 g^2_X+ \hat{y}^2_e \right).
\end{eqnarray}
Here we have ignored the contributions from the Yukawa interactions involving the heavy crossing scalar $\sigma$. As an example, we shall consider a quasi-degenerate neutrino spectrum, i.e. $\delta\hat{y}^2_{\nu}=\hat{y}^2_{\nu}-\bar{y}_{3}^{2}\ll \bar{y}^2_{1,2,3}$, and take $3g_R^2/4+g_X^2/4+3\bar{y}_3^{2}/4> \hat{y}_e^{2}/2$. Under these assumptions, we can perform
 \begin{eqnarray}
\Omega_{l'^{}_R}^{}\!\!&\simeq&\!\! \Omega_{l'^{}_R}^{(0)}+\Omega_{l'^{}_R}^{(1)}\,,~\Omega_{l'^{}_R}^{(0)}\!\!\simeq \!\!\frac{T}{2\sqrt{2}}\sqrt{\frac{3}{4}g_R^2+\frac{1}{4}g_X^2+\frac{3}{4}\bar{y}^{2}_3}\,,\nonumber\\
\Omega_{l'^{}_R}^{(1)}\!\!&\simeq&\!\!\frac{T}{4\sqrt{2}}\frac{\left(\frac{3}{4}U^\ast_{}\delta\hat{y}^2_{\nu} U^T_{} +\frac{1}{2}\hat{y}^2_e\right)}{\sqrt{\frac{3}{4}g_R^2+\frac{1}{4}g_X^2+\frac{3}{4}\bar{y}^{2}_3}}\,,~~\Omega_{e'^{}_L}^{}\!\!=\!\!\frac{T}{2\sqrt{2}}\sqrt{g^2_X+ \hat{y}^2_e}\,,\nonumber\\
\Omega_0^{} \!\!&=&\!\!\frac{\Omega_{l'^{}_R}^{(0)} + \Omega_{e'^{}_L}^{}}{2}\,,~~\delta p\!=\!\!-3 \Omega_{l'^{}_R}^{(1)}\,,~\Delta k \!=3\!\left(\Omega_{l'^{}_R}^{(0)} - \Omega_{e'^{}_L}^{}\right), \nonumber\\
\bar{\gamma}\!\!&=&\!\!\frac{\gamma_{l'^{}_R}^{}+ \gamma_{e'^{}_L}^{}}{2}\sim \frac{T}{64\pi}\left(9g_R^2+18 g_X^2+\frac{3}{4}\bar{y}_3^2\right).
\end{eqnarray}
When the perturbation condition  $2\Omega_{l'^{}_R,e'^{}_L}^{}>  \hat{y}_e^{} v'^{}_{1}$ is satisfied, the reflection asymmetry $\Delta (\omega)$ can be analytically solved by \cite{hs1995}
\begin{eqnarray}
\label{formula}
\!\!\!\!\Delta (\omega)\!\!&=&\!\!-i\frac{\textrm{Tr}\left[(\hat{y}_e^{} v'^{}_1)^2_{}, \delta p \right]^3_{}}{2^7_{}\cdot 3^{10}_{}\bar{\gamma}^{9}_{}}\left[\!1+\!\left(\frac{\omega -\Omega_0^{}}{\bar{\gamma}}\right)^2_{}\right]^{-6}_{}\nonumber\\
\!\!\!\!\!\!&=&\!\!\frac{3^3_{} 2^{\frac{57}{2}}_{}\pi^9_{} \bar{y}_{\tau}^6 \bar{y}_3^6 \left[\!1+\!\left(\frac{\omega -\Omega_0^{}}{\bar{\gamma}}\right)^2_{}\right]^{\!-6}_{}}{\left(\frac{3}{4}g^{2}_{R}+\frac{1}{4}g^2_{X}+\frac{3}{4}\bar{y}_3^2\right)^{\!\frac{3}{2}}_{}\!\!\left(9g_R^2+18 g_X^2+\frac{3}{4}\bar{y}_3^2\right)^{\!9}_{}} \left(\!\frac{v'^{}_{1}}{T}\!\right)^{\!6}_{}\nonumber\\
\!\!\!\!\!\!\!\!&&\!\!\times \!\!\prod_{i>j \atop e,\mu,\tau}\!\!\!\left(\frac{m_{i}^{2}-m_{j}^{2}}{m_\tau^2}\right)\!\!\prod_{i>j \atop 1,2,3}^{}\!\!\!\left(\frac{m_{\nu_i^{}}^2-m_{\nu_j^{}}^2}{m_{\nu_3^{}}^2}\right) J_{CP}^{}\,,
\end{eqnarray}
with $J_{CP}^{}=\frac{1}{4} \sin 2\theta_{12}^{}\sin 2 \theta_{23}^{}\cos^2_{}\theta_{13}^{}\sin\theta_{13}^{}\sin\delta$. Clearly, the above formula depends on the product $(m_{\nu_3^{}}^2-m_{\nu_1^{}}^2)(m_{\nu_3^{}}^2-m_{\nu_2^{}}^2)$ so that it will not be sensitive to the normal or inverted hierarchy of the neutrino mass matrix. For the known parameters  $\Delta m_{21}^2=m_2^2-m_1^2=7.37\times 10^{-5}_{}\,\textrm{eV}^2_{}$, $\Delta m_{31}^2(\Delta m_{23}^2)=m_3^2- m_1^2(m_2^2-m_3^2)=2.56(2.54)\times 10^{-3}_{}\,\textrm{eV}^2_{}$, $\sin ^2_{}\theta_{12}^{}=0.297$, $\sin ^2_{}\theta_{23}^{}=0.425(0.589)$, $\sin ^2_{}\theta_{13}^{}=0.0215(0.0216)$, $m_\tau^{}=1.78\,\textrm{GeV}$, $m_\mu^{}=106\,\textrm{MeV}$, $m_e^{}=511\,\textrm{keV}$, $m_b^{}=4.17\,\textrm{GeV}$, $g_R^{}=0.653$, and $g_X^{}=0.428$ \cite{pdg2018}, we can input $\bar{f}_3^{}=\sqrt{4\pi}$, $\bar{y}_{\tau}^{}=\sqrt{4\pi}m_\tau^{}/m_b^{}=1.51$ to obtain $\Omega_0^{}\sim 0.629\,\textrm{T}$, $\bar{\gamma}\sim 0.0824\,\textrm{T}$, $\Delta k\sim 2.86\,\textrm{T}$. The perturbation condition $2\Omega_{l'^{}_R,e'^{}_L}^{}>  \hat{y}_e^{} v'^{}_{1}$ then can constrain $v'^{}_1/T \lesssim 0.675$. We eventually can estimate an upper bound on the mirror baryon and lepton numbers, 
\begin{eqnarray}
\label{lnumber2}
\frac{n_{L'}^{}}{s}&=&\frac{n^{}_{B'}}{s}\sim 1.3 \times 10^{-10}_{}\times \left(\frac{\kappa}{20}\right)\!\left(\frac{c_L^{}}{1}\right)\!\left(\frac{0.1}{v_W^{}}\right)\nonumber\\
&&\times \!\left(\frac{v'^{}_{1}/T}{0.675}\right)^{\!\!6}_{}\! \left(\frac{0.1\,\textrm{eV}}{m_{\nu_3^{}}^{}}\right)^{\!\!6}_{}\!\left(\frac{\sin\delta}{0.01}\right)\,.
\end{eqnarray}
It should be noted that the result (\ref{lnumber2}) is based on the assumptions, $\delta\hat{y}^2_{\nu}=\hat{y}^2_{\nu}-\bar{y}_3^{2}\ll \bar{y}^2_{1,2,3}$, $3g_R^2/4+g_X^2/4+3\bar{y}_3^{2}/4> \hat{y}_e^{2}/2$ and $2\Omega_{l'^{}_R,e'^{}_L}^{}> \hat{y}_e^{} v'^{}_{1}$. Clearly, this is quite a rough estimation. For a more reliable result, it should be necessary to do a numerical calculation. For simplicity we do not study the numerical calculation here.

\section{Ordinary baryon number}

The heavy crossing scalar $\sigma$ will mediate the three-body decays of a mirror charged fermion $f'$ into an ordinary fermion $f$ and two dark matter scalars $\chi$. The gauge boson $W^{\pm}_{R}$ will mediate the three-body decays of one mirror fermion into three mirror fermions. We require the second and third generations of mirror charged fermions can not efficiently decay into the ordinary fermions with the dark matter scalar, instead, their decays will be dominated by the gauge interactions. Furthermore, the decays of the first-generation mirror charged fermions into their ordinary partners with the dark matter scalar should go into equilibrium before the BBN. These requirements can be achieved by choosing the related masses and Yukawa couplings. 

In our model, the lepton number is explicitly broken by the cubic couplings of the Higgs triplets to the Higgs doublets. The Higgs triplets thus can acquire their seesaw-suppressed VEVs. In consequence, the ordinary and mirror neutrinos can independently obtain their Majorana masses through the related Yukawa couplings. It should be noted the mirror neutrinos have no mixing with the ordinary neutrinos because of the unbroken $Z_4^{}\times Z_4^{}$ symmetry. The mirror neutrinos can be allowed very light if their decoupled temperatures are above the ordinary QCD scale. The interactions for generating the Majorana neutrino masses will lead to some lepton-number-violating processes and will wash out any lepton numbers. These processes potentially can wash out any baryon numbers when the baryon and lepton numbers are connected by the sphaleron processes. Fortunately, the lepton-number-violating interactions for generating the desired neutrino masses can keep departure from equilibrium before the ordinary sphaleron processes stop working \cite{ghsz2009,gm2017}.

We eventually can obtain a nonzero ordinary baryon asymmetry $B$ from the mirror baryon asymmetry $B'$ and/or the mirror lepton asymmetry $L'$ $(B'\equiv L')$ in the following ways:
\begin{itemize}
\item If the mirror quarks as well as the mirror leptons decay into the ordinary fermions after the electroweak phase transition, none of the mirror baryon and lepton asymmetries will participate in the ordinary sphaleron processes. The mirror baryon asymmetry should be exactly equal to the ordinary baryon asymmetry,
\begin{eqnarray}
B=B'=L'\,.
\end{eqnarray}
\item if the mirror quarks rather than the mirror leptons decay into the ordinary fermions before the electroweak phase transition, all of the mirror baryon asymmetry will participate in the ordinary sphaleron processes. Thus the mirror baryon asymmetry will be partially converted to the ordinary baryon asymmetry \cite{ht1990},
\begin{eqnarray}
B=\frac{28}{79}B'=\frac{28}{79}L'\,.
\end{eqnarray}
\item if the mirror leptons rather than the mirror quarks decay into the ordinary fermions before the electroweak phase transition, all of the mirror lepton asymmetry will participate in the ordinary sphaleron processes. Thus the mirror lepton asymmetry will be partially converted to the ordinary baryon asymmetry \cite{ht1990}, 
\begin{eqnarray}
B=B'-\frac{28}{79}L'= \frac{51}{79}L'\,.
\end{eqnarray}
\item if either the mirror quarks or the mirror leptons decay into the ordinary fermions during the electroweak phase transition, a nonzero difference between the mirror baryon and lepton asymmetries will participate in the ordinary sphalerons. Thus the ordinary baryon asymmetry can be given by \cite{ht1990}
\begin{eqnarray}
B_q^{}&=&\left(1-r_{q'}^{}\right)B'+\frac{28}{79}\left(r_{q'}^{}B'-r_{l'}^{}L'\right)\nonumber\\
&=&\left(1-\frac{51}{79}r_{q'}^{}-\frac{28}{79}r_{l'}^{}\right)L'\nonumber\\
&&\textrm{with}~~r_{q',l'}^{}\leq 1\,,~~r_{q'}^{}\neq r_{l'}^{}\,.
\end{eqnarray}
\end{itemize}

\section{Conclusion}

In this paper we have demonstrated a novel scenario for explaining the cosmic baryon asymmetry. In our scenario, during a strongly first order phase transition of the spontaneous left-right symmetry breaking, the CP-violating reflections of the mirror fermions off the mirror Higgs bubbles can generate a mirror lepton asymmetry and an equal mirror baryon asymmetry. An ordinary baryon asymmetry then can be induced by the three-body decays of a mirror fermion into an ordinary fermion and two dark matter scalars. Benefitted from a parity symmetry for solving the strong CP problem, the cosmic baryon asymmetry can be well described by the ordinary lepton mass matrices up to an overall factor. In this scenario, the Dirac CP phase in the Majorana neutrino mass matrix provides a unique source for the required CP violation. Our scenario thus can be ruled out in the future if the Dirac CP phase has a too small value. Furthermore, the Higgs triplet for type-II seesaw as well as the first-generation mirror charged fermions are allowed at the TeV scale so that they may be verified at the colliders such as the LHC \cite{fhhlw2008,cgnr2013}. The dark matter scalar can be connected to the TeV-scale Higgs triplet and/or the SM Higgs scalar \cite{ht2016,los2017}.

\textbf{Acknowledgement}: This work was supported by the National Natural Science Foundation of China under Grant No. 11675100 and the Recruitment Program for Young Professionals under Grant No. 15Z127060004.

\end{document}